\documentclass[conference]{IEEEtran}
\IEEEoverridecommandlockouts

\usepackage{cite}
\usepackage{amsmath,amssymb,amsfonts}
\usepackage{algorithmic}
\usepackage{graphicx}
\usepackage{textcomp}
\usepackage{float}
\usepackage{multirow}
\usepackage{xcolor}
\usepackage{booktabs} 

\def\BibTeX{{\rm B\kern-.05em{\sc i\kern-.025em b}\kern-.08em
    T\kern-.1667em\lower.7ex\hbox{E}\kern-.125emX}}
\begin{document}

\title{Towards Low-Latency and Adaptive Ransomware Detection Using Contrastive Learning}

\author{\IEEEauthorblockN{1\textsuperscript{st} Zhixin Pan}
 \IEEEauthorblockA{\textit{College of Engineering} \\
 \textit{Florida State University}\\
 Tallahassee, USA \\
 zp23e@fsu.edu}
 \and
 \IEEEauthorblockN{2\textsuperscript{nd} Ziyu Shu}
 \IEEEauthorblockA{\textit{Department of Radiation Oncology} \\
 \textit{Stony Brook University}\\
 Stony Brook, USA \\
 zs919@nyu.edu}
 \and
 \IEEEauthorblockN{3\textsuperscript{rd} Amberbir Alemayoh}
 \IEEEauthorblockA{\textit{College of Engineering} \\
 \textit{Florida State University}\\
 Tallahassee, USA \\
amberbirmehabaw.alemayoh@studenti.univr.it}
 }
 

\maketitle
\let\thefootnote\relax\footnotetext{This paper was published in the 2025 IEEE International Conference on Computer Design}
\begin{abstract}
Ransomware has become a critical threat to cybersecurity due to its rapid evolution, the necessity for early detection, and growing diversity, posing significant challenges to traditional detection methods. While AI-based approaches had been proposed by prior works to assist ransomware detection, existing methods suffer from three major limitations, ad-hoc feature dependencies, delayed response, and limited adaptability to unseen variants. In this paper, we propose a framework that integrates self-supervised contrastive learning with neural architecture search (NAS) to address these challenges. Specifically, this paper offers three important contributions.
(1) We design a contrastive learning framework that incorporates hardware performance counters (HPC) to analyze the runtime behavior of target ransomware.
(2) We introduce a customized loss function that encourages early-stage detection of malicious activity, and significantly reduces the detection latency.
(3) We deploy a neural architecture search (NAS) framework to automatically construct adaptive model architectures, allowing the detector to flexibly align with unseen ransomware variants.
Experimental results show that our proposed method achieves significant improvements in both detection accuracy (up to 16.1\%) and response time (up to 6x) compared to existing approaches while maintaining robustness under evasive attacks.
\end{abstract}

\begin{IEEEkeywords}
Machine Learning, Ransomware, Security
\end{IEEEkeywords}

\section{Introduction}
\label{sec:intro}

Ransomware has emerged as one of the most pervasive threats in cybersecurity. It encrypts files on infected machines and demands a ransom for decryption, resulting in significant financial losses. According to a recent study~\cite{ispahany2024ransomware}, global ransomware-related damages have exceeded \$6 trillion, highlighting an urgent need for efficient defense frameworks.
 
\begin{figure}[htbp]
\centering
\vspace{-0.2 in}
\includegraphics[width=0.48\textwidth]{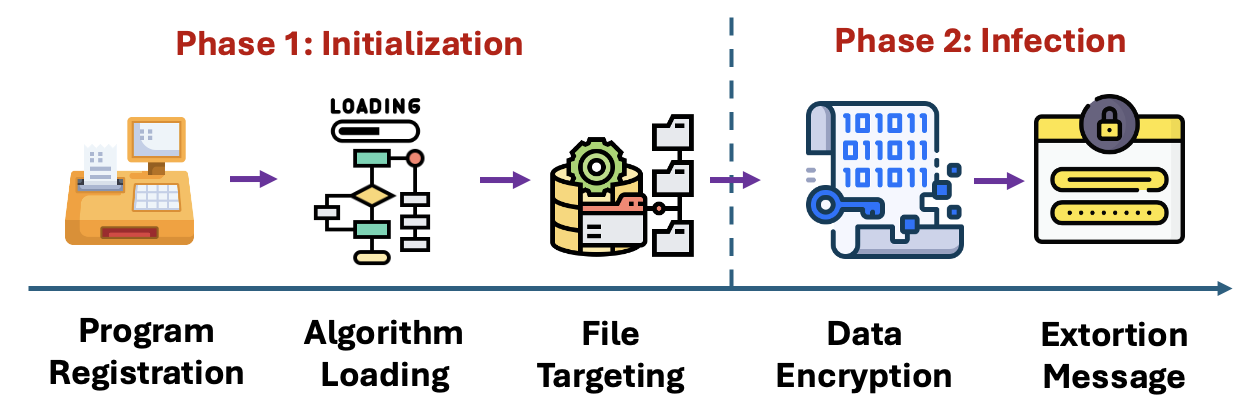}
\vspace{-0.2 in}
\caption{Illustration of a typical ransomware infection workflow. The attack begins with a stealthy initialization phase for registering for persistence and encryption algorithm loading, followed by the infection phase with data encryption and extortion message displaying.}
\vspace{-0.2 in}
\label{fig:timeline}
\end{figure}

Compared with conventional malware, ransomware poses a greater threat due to its stealth and urgency for immediate response. As illustrated in Figure~\ref{fig:timeline}, a typical ransomware attack involves two major phases: a stealthy initialization phase where the malware registers itself and loads encryption algorithms, along with the infection phase where encryption begins and causes damage within milliseconds. The initialization phase typically possesses similar behavior to benign programs, making early detection particularly difficult. Meanwhile, the encryption phase progresses extremely quickly; even if detected and terminated, the ransomware may have already encrypted critical files, causing irreversible damage. Furthermore, modern ransomware continues to evolve through obfuscation, code morphing, and logic camouflage, producing sophisticated variants capable of evading traditional detectors.

Traditional detection methods can be broadly categorized into two categories: static and dynamic analysis. Static analysis inspects executable files before runtime and identifies ransomware utilizing rule-based identification or signature matching. While computationally efficient, static methods are vulnerable to evasive attacks with code morphing or insertion of non-functional blocks to disrupt signature patterns~\cite{ispahany2024ransomware}, leading to poor security reliability. In contrast, dynamic analysis monitors runtime behaviors such as unusual file access, memory activity, or register values. Although dynamic analysis provides richer behavioral contexts, it often suffers from detection latency, which is often unacceptable given the characteristics of ransomware attacks.

Recent studies in the field have turned interest to Machine learning (ML) based ransomware detection approaches, which are capable of learning complex behavioral patterns efficiently, obtaining promising detection performance. However, several major limitations remain unaddressed. First, they heavily rely on manually selected features, limiting their generalizability and robustness to evasion attacks. Second, most models are still trained for accuracy alone and do not explicitly penalize detection delays, reducing their real-world responsiveness. Moreover, the architectures used are often statically designed, limiting their adaptability to unseen ransomware variants.

In this paper, we propose a novel framework that integrates contrastive learning and neural architecture search (NAS) to achieve low-latency and adaptive ransomware detection. Specifically, this paper makes the following key contributions:

\begin{itemize}
    \item \textbf{Contrastive learning:} We design a contrastive learning framework achieving automated feature engineering. It exempts us from the ad-hoc feature selection, while also improving the resilience towards evasive attacks.
    
    \item \textbf{Latency-aware detection loss:} We introduce a custom training objective that encourages earlier detection of ransomware, significantly reducing the detection latency.
    
    \item \textbf{Neural architecture search:} We employ a neural architecture search framework to automatically discover expressive model structures tailored for target tasks while maintaining flexibility to be adapted to unseen variants.

\end{itemize}

The rest of the paper is organized as follows: Section~\ref{sec:bgd} reviews related work. Section~\ref{sec:prop} details the proposed detection framework. Experimental results are reported in Section~\ref{sec:exp}. We conclude in Section~\ref{sec:conc}.
\section{Related Works and Background}
\label{sec:bgd}

\subsection{ML-based Ransomware Detection}

As discussed in Section~\ref{sec:intro}, the increasing sophistication of ransomware attacks has prompted the adoption of machine learning (ML) approaches in this field. Similar to conventional defense strategies, existing ML-based ransomware detection methods can also be broadly categorized into static and dynamic analysis.

\textbf{Static analysis} focuses on inspecting ransomware source code or executable files before execution to identify suspicious signatures using ML-based anomaly detection. For instance, Lee et al.~\cite{lee2025ml} proposed a multilayer perceptron (MLP) based method~\cite{seiffert2001multiple} to detect ransomware signatures from file headers. Similar approaches have been explored using other ML models, such as k-nearest neighbors (KNN)~\cite{larocque2024effective}, deep neural networks (DNN)~\cite{marcany2024innovative}, and reinforcement learning~\cite{adamov2020reinforcement}. However, static analysis methods often suffer from high false positive rates, as benign programs frequently exhibit patterns similar to ransomware (e.g., disk encryption utilities), leading to high false positive rates. Furthermore, sophisticated adversaries can adopt obfuscation techniques such as code morphing to bypass naive signature-based detectors.

\textbf{Dynamic analysis}, in contrast, monitors program behavior during execution to identify malicious actions. For example, Herrera et al.~\cite{herrera2023dynamic} developed a dynamic ransomware detection framework that uses random forests to recognize suspicious activity based on manually selected features. Similarly, Gulmez et al.~\cite{gulmez2024xran} developed XRan, an explainable deep learning-based ransomware detection system to analyze temporal patterns for improved detection accuracy. While dynamic analysis offers promising detection performance, most of these methods fail to take the detection latency into consideration, resulting in irreversible damage even after successful detection.

As we can see from the above discussion, despite the developments of ML-based detection methods, challenges still persist. First, many existing approaches rely on manually engineered features tailored to specific ransomware types, limiting their adaptability to new or evolving variants. Second, most dynamic analysis methods are optimized for accuracy only, neglecting the importance of early detection to mitigate the damage that would occur during execution. Finally, current models rely on fixed architectures, which hinders their adaptivity to rapidly evolving ransomware families.

These limitations underscore the need for more robust and adaptive detection frameworks. Our proposed approach aims to address these challenges utilizing a self-supervised contrastive learning algorithm, introduced in the next subsection.

\subsection{Contrastive Learning}

The limitations of existing ML-based detection methods, particularly their reliance on manually selected features, have highlighted the need for more adaptive approaches. A promising direction to address this challenge is through \textit{contrastive learning}, which has demonstrated remarkable success in automated feature extraction.

Contrastive learning is a special type of \textit{ self-supervised learning} (SSL). It focuses on training an encoder to extract meaningful representations by learning to distinguish between similar and dissimilar data samples. Specifically, it maps inputs into a feature space where similar sample pairs are pulled closer together, while dissimilar pairs are pushed apart. This enables the model to capture essential patterns automatically, without manual feature engineering.

\begin{figure}[htp]
    \vspace{-0.1in}
    \centering
    \includegraphics[width=0.75\linewidth]{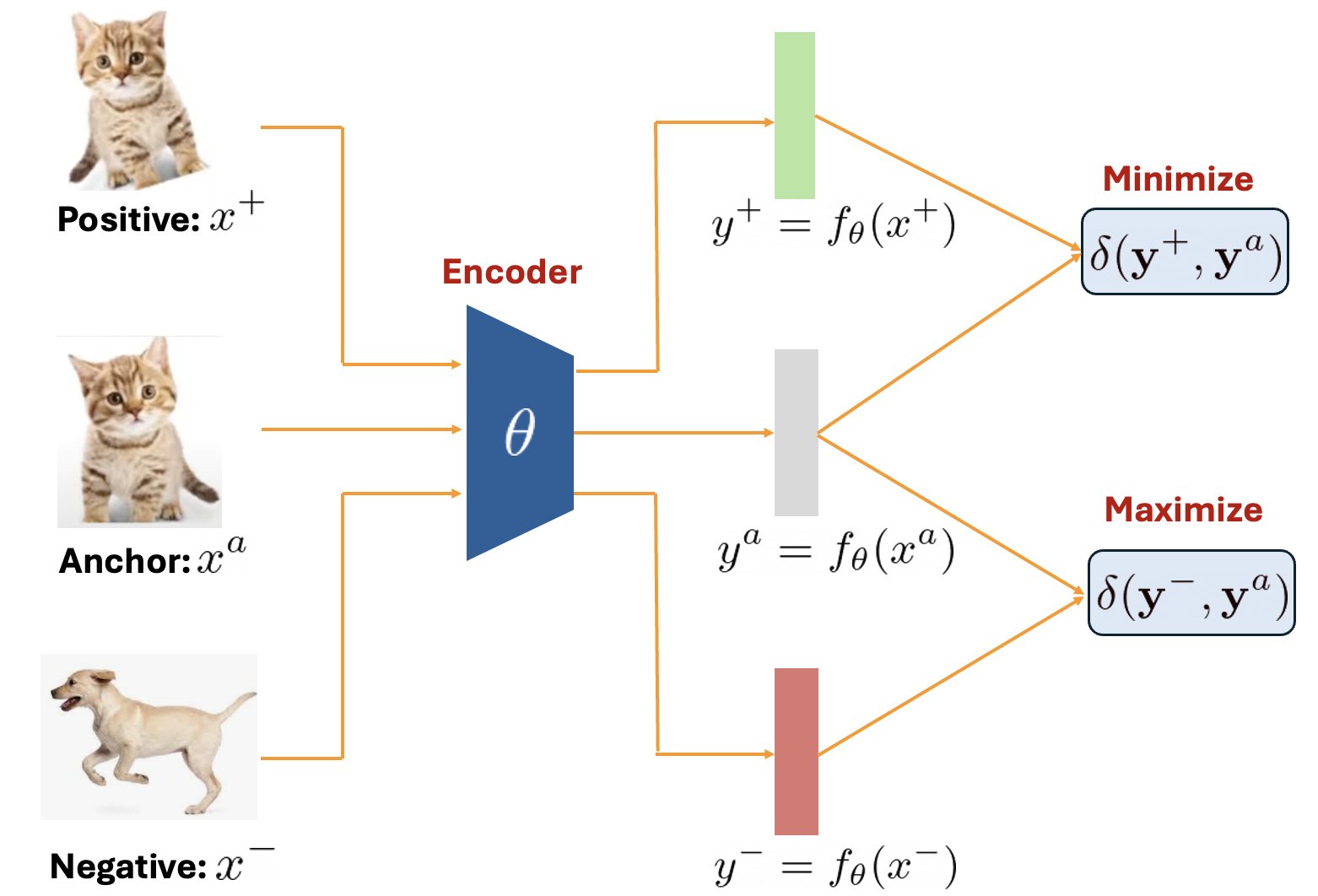}
    \caption{Illustration of contrastive learning. Given an anchor input \( x^a \), a positive example \( x^+ \) is generated through data augmentation, while a negative example \( x^- \) is selected from a different class. The model learns a feature representation such that the distance \( \delta(x^a, x^+) \) is minimized, while the distance \( \delta(x^a, x^-) \) is maximized.}
    \vspace{-0.1in}
    \label{fig:contrastive_learning}
\end{figure}

Figure~\ref{fig:contrastive_learning} presents the basic contrastive learning workflow, commonly applied in computer vision. For a given input (referred to as the \emph{anchor}), a \emph{positive} sample is generated by applying data augmentations (e.g., random cropping, rotation, or color jittering), ensuring high semantic similarity with the anchor. In contrast, a \emph{negative} sample is selected from a different class, representing dissimilar content. All samples are passed through a shared encoder model, which extracts their feature embeddings, followed by a distance-measuring function. The model is then trained to minimize the distance between the anchor and its positive pair while maximizing that to negative samples. This learning objective encourages the model to extract discriminative and meaningful feature embeddings automatically.

Compared to traditional supervised learning methods, contrastive learning does not require manually defined features or exhaustive labeled datasets. Instead, it leverages inherent relationships within the data, reducing the dependence on expert-crafted features. A prior study has also applied contrastive learning for ransomware detection, proposed in~\cite{yang2022android}. However, its design was tightly coupled with Android-specific features such as system call patterns and mobile app behaviors, limiting its generalizability to other platforms. 


\section{Proposed Method}
\label{sec:prop}

\subsection{Overview}
\label{sec:overview}
To motivate our proposed method, we summarize the key limitations of existing techniques discussed in Section~\ref{sec:bgd}:

\begin{itemize}
    \item \textbf{Ad-hoc feature engineering:} Many existing methods rely on manually crafted features, making them vulnerable to evasion techniques such as code obfuscation.
    
    \item \textbf{Delayed detection response:} Most dynamic analysis models are optimized solely for accuracy, without explicitly addressing the need for detection latency, which is critical in preventing ransomware-induced damage.
    
    \item \textbf{Limited adaptability to new variants:} Prior approaches often adopt static model architectures that lack the flexibility to adapt to emerging ransomware families.
\end{itemize}

\begin{figure*}[htp]
    \centering
    \includegraphics[width=0.87\linewidth]{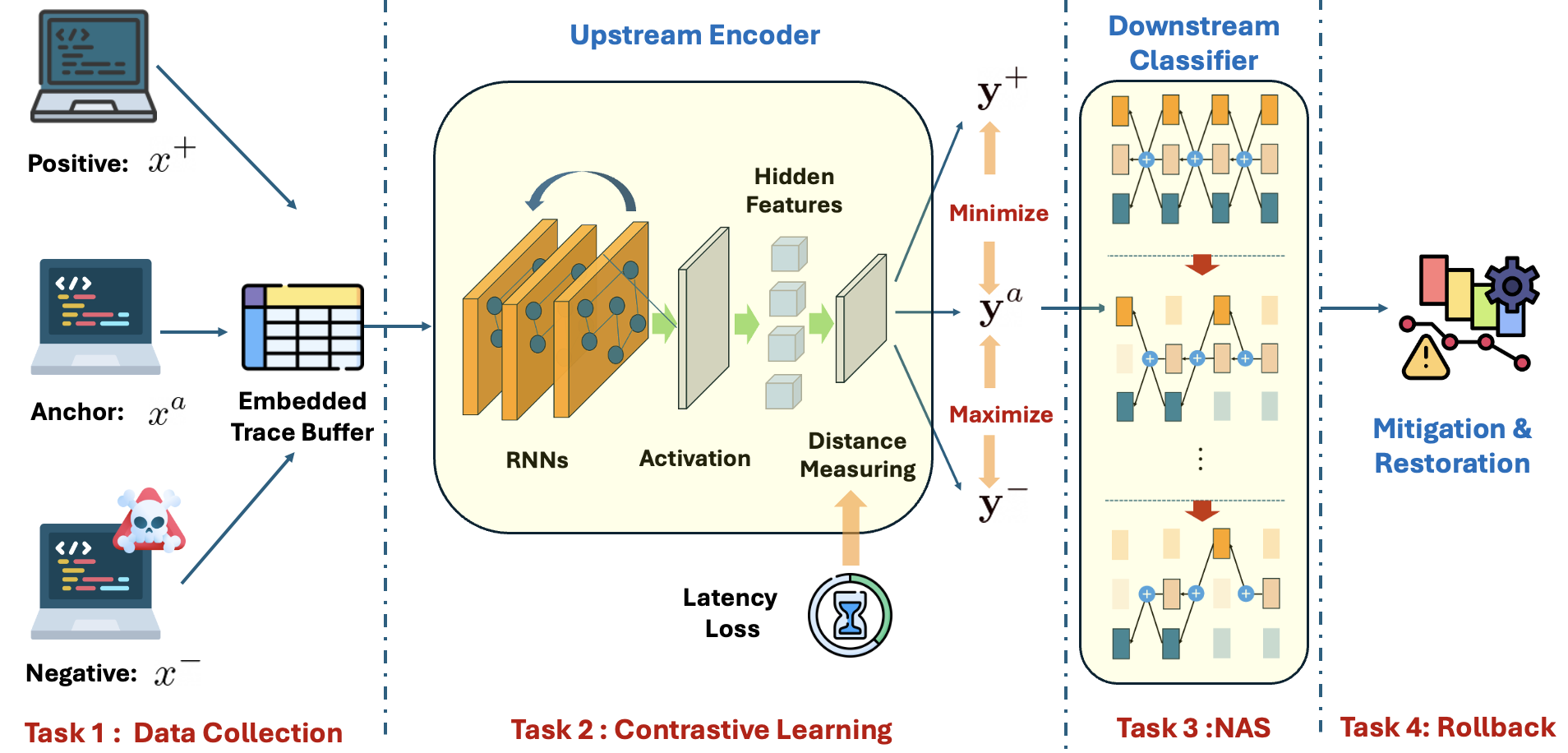}
    \caption{Overview of the proposed ransomware detection framework. The system consists of four major components: (1) Hardware-assisted data collection uses Embedded Trace Buffers (ETBs) to capture fine-grained, real-time execution traces; (2) Contrastive learning-based upstream encoder processes windowed trace sequences to extract hidden features, followed by a distance measuring step. The learning process adopts a latency-aware loss promoting early detection; (3) The NAS-guided downstream classifier dynamically adapts architecture for improved generalizability, and (4) The detection and rollback module monitors runtime behavior and performs data recovery upon detection.}
    \vspace{-0.15 in}
    \label{fig:overview}
\end{figure*}

To address these challenges, we propose a fully automated learning framework that integrates contrastive self-supervised learning, a latency-aware training objective, and neural architecture search (NAS) for model optimization. Figure~\ref{fig:overview} illustrates the overview of the proposed framework, which consists of an \textit{upstream encoder} and a \textit{downstream classifier}, and the workflow can be decomposed into four major tasks:

\begin{itemize}
    \item \textbf{Hardware-assisted data collection:} We adopt dynamic analysis by leveraging hardware components to collect time-sequential traces of runtime behavior. These traces are segmented into windowed sequences for compatibility with sequential learning models. (Section~\ref{sec:preprocess})
    
    \item \textbf{Contrastive learning-based upstream encoder:} A contrastive learning framework is used to train an RNN-based encoder that automatically extracts feature representations from sequential trace data. A latency-aware loss function is further incorporated to prioritize early detection. (Section~\ref{sec:ssl})
    
    \item \textbf{NAS-guided downstream classifier:} A classifier is trained atop the encoder outputs to make final ransomware predictions. To enhance its adaptivity, we adopt a NAS framework that automatically explores optimal architectures, while allowing it to better align with unseen ransomware variants. (Section~\ref{sec:nas})

    \item \textbf{Real-time detection and rollback:} During deployment, the model continuously monitors input traces and raises alerts upon detection of malicious activity. It then performs system isolation and data restoration to mitigate damage. (Section~\ref{sec:mitigation})
\end{itemize}

\subsection{Hardware-assisted data collection}
\label{sec:preprocess}
Given the limitations of static analysis, particularly its vulnerability to obfuscation and inability to reflect runtime behavior, we adopt dynamic analysis as the foundation of our proposed framework. However, as discussed in Section~\ref{sec:bgd}, dynamic analysis often suffers from detection latency. To address this, we apply two key strategies. First, instead of relying on software instrumentation (which introduces latency and potential noise), we leverage hardware-assisted data collection using Embedded Trace Buffers (ETBs) to unobtrusively monitor real-time program execution. Second, we incorporate a latency-aware loss function during model training to encourage early-stage detection (see Section~\ref{sec:ssl} for details).

The raw traces captured by ETBs consist of sequential buffer values that log control flow transitions, memory access patterns, and low-level instruction behavior. These fine-grained signals are critical for identifying phase transitions and anomalous encryption activities during ransomware infection, which has been validated in~\cite{pan2022hardware}. Notice that these traces are directly captured from raw program execution and reflect intrinsic behavioral properties rather than manually engineered features. Our approach does not rely on handcrafted feature selection but instead leverages fundamental runtime characteristics to construct the sequential input data to ML models.

\begin{figure}[H]
    \vspace{-0.1in}
    \centering
    \includegraphics[width=\linewidth]{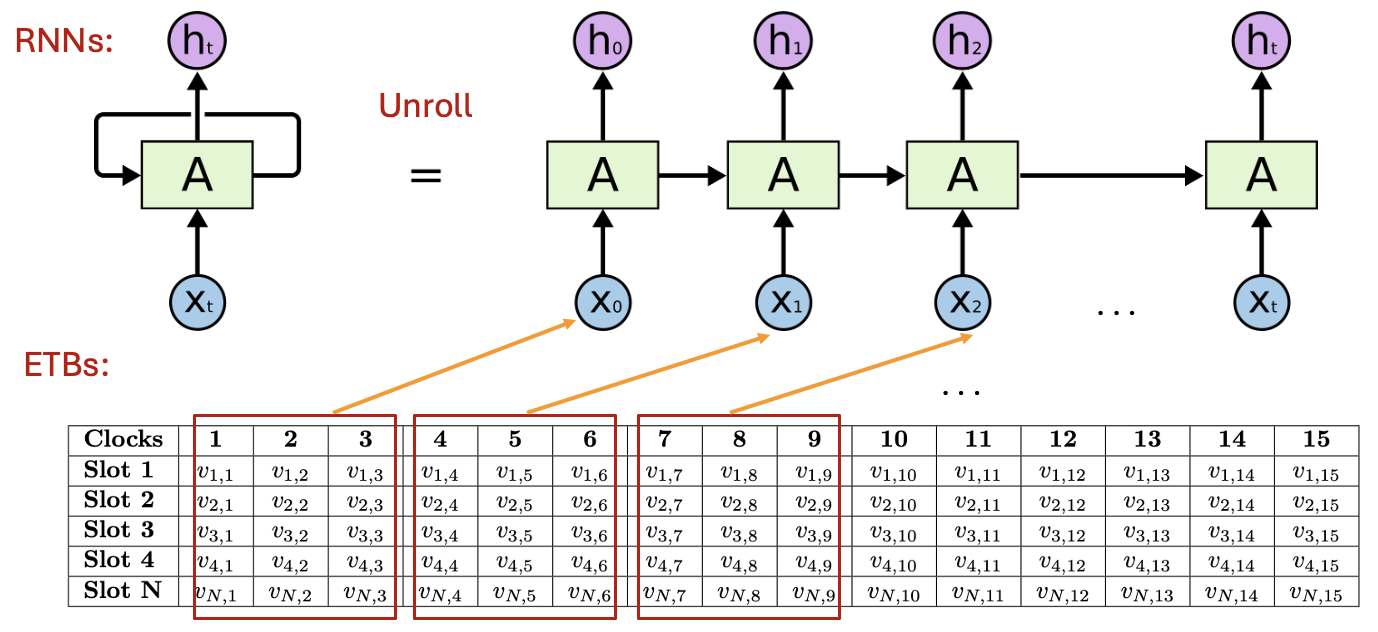}
    \caption{Illustration of the hardware-assisted trace windowing process. Execution traces collected via Embedded Trace Buffers (ETBs) are organized as a matrix, where rows correspond to different buffer slots and columns represent clock cycles. The trace stream is segmented into fixed-size sliding windows, each representing a short temporal sequence \( x_i \). These windows are then fed sequentially into a recurrent neural network (RNN), which encodes each input window into a corresponding hidden representation \( h_i \).}
    \vspace{-0.1in}
    \label{fig:tracewindow}
\end{figure}

To handle the time-sequential traces, we segment the continuous stream into fixed-size sliding windows. Each window encapsulates a short activity segment while preserving temporal structure. This windowing mechanism ensures compatibility with sequential models such as recurrent neural networks (RNNs) and supports both real-time and batch-based processing. The synergy between ETBs and RNNs is illustrated in Figure~\ref{fig:tracewindow}. Note that by focusing on runtime behavior rather than static code patterns, our framework inherently resists common evasion techniques such as code morphing, which typically target static analysis tools by altering program structure without affecting execution semantics.


\subsection{Contrastive learning based upstream encoder}
\label{sec:ssl}

\noindent \underline{\textbf{Architecture:}}
To process the ETB-derived sequences introduced in Section~\ref{sec:preprocess}, we utilize the recurrent neural network (RNN). Specifically, we employ a three-layer gated recurrent unit (GRU)~\cite{dey2017gate}, a lightweight RNN variant to extract meaningful representations from the time-sequential data.

Each input trace \( x_i = \{x_i^1, x_i^2, ..., x_i^T\} \) is fed into the GRU sequentially, producing a corresponding sequence of hidden states \( h_i = \{h_i^1, h_i^2, ..., h_i^T\} \), where the output length may vary depending on the input. These hidden states capture the temporal evolution of program behavior and serve as the latent representation of the input. The resulting embeddings are then used to compute similarity scores between input pairs in the contrastive learning framework.

\noindent \underline{\textbf{Distance Function:}}
Measuring the similarity between sequential traces in ransomware detection poses several challenges. First, due to variations in program execution time, the ETB-derived input traces often differ in length, making conventional distance metrics like Euclidean distance unsuitable. Second, our framework is designed for online detection, i.e., it receives runtime data incrementally and must evaluate similarity in a streaming fashion. Therefore, the distance function must support partial alignment and progressive updates. Finally, even for the same ransomware variant, attackers may inject {non-functional instructions or reorder operations} to obfuscate behavior, resulting in shifted patterns in the temporal domain. Under such circumstances, a similarity measurement should still detect malicious behavior by aligning semantically equivalent segments that are temporally displaced. To address these challenges, we adopt \textit{Dynamic Time Warping (DTW)} as our core distance metric.

\begin{figure}[htp]
    \vspace{-0.1in}
    \centering
    \includegraphics[width=\linewidth]{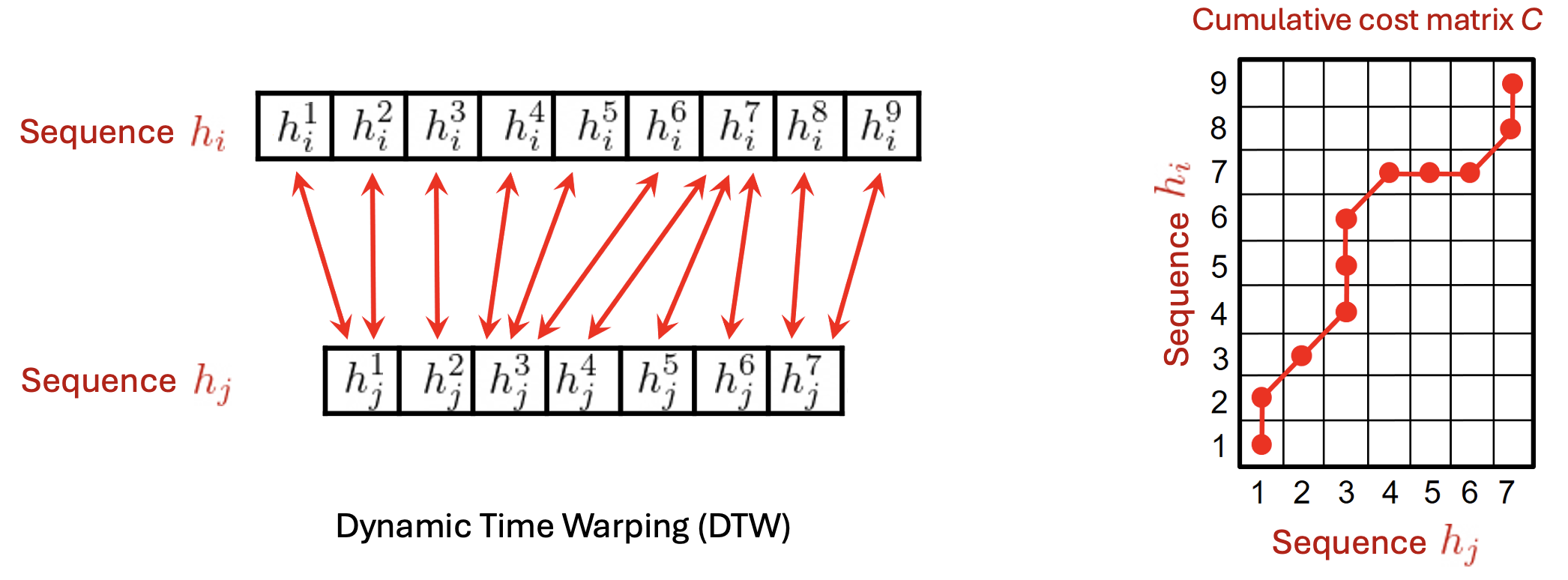}
    \caption{Illustration of the DTW algorithm (Image Credit:~\cite{muller2021fundamentals}). The optimal path with minimum cumulative distance is shown in the right panel of the figure, illustrating how the result was obtained through the DP recurrence. Accordingly, each of the red bidirectional arrows in the left panel encodes the local correspondence between elements guided by the optimal path.}
    \vspace{-0.1in}
    \label{fig:dtw}
\end{figure}

DTW is a dynamic programming (DP) based algorithm that measures the similarity between two temporal sequences by finding an optimal alignment that minimizes the cumulative difference. Given two hidden sequences, \( h_i = \{h_i^1, ..., h_i^{T_i}\} \) and \( h_j = \{h_j^1, ..., h_j^{T_j}\} \), DTW first computes a cost matrix \( D \in \mathbb{R}^{T_i \times T_j} \), where each element \( D(p, q) \) represents the squared distance between \( h_i^p \) and \( h_j^q \). A cumulative cost matrix \( C \in \mathbb{R}^{T_i \times T_j} \) is then computed using the following recurrence equation:
\vspace{-0.1 in}
\begin{equation}
C(p, q) = D(p, q) + \min \begin{cases}
C(p-1, q), \\
C(p, q-1), \\
C(p-1, q-1)
\end{cases}
\end{equation}
The distance between \( h_i \) and \( h_j \) is finally given by the cumulative cost along the optimal path inside $C$ that minimizes $C(T_i,T_j)$. Figure~\ref{fig:dtw} depicts an illustrative example. Intuitively, this minimum cumulative distance captures the lowest cost of making adjustments to match similar patterns between two sequences, despite temporal distortions. Based on this, we define the distance between \( h_i \) and \( h_j \) as \begin{equation}
    d(h_i,h_j) = \frac{1}{2}C^2(T_i,T_j)
\end{equation}

As we can see, DTW effectively addresses the challenges discussed above. First, it naturally accommodates input sequences of varying lengths. Second, as a DP-based algorithm, DTW incrementally updates its cost matrix, enabling online computation. Finally, observe that a single element in the shorter sequence \( h_j \) can be aligned with multiple elements in the longer sequence \( h_i \); this demonstrates DTW's robustness to temporal distortions. For example, if \( h_j \) represents a vanilla ransomware trace and \( h_i \) corresponds to an obfuscated variant with injected delays, DTW can still successfully align its core behavior patterns despite timing misalignments.

\noindent \underline{\textbf{Training Loss:}}  
To effectively train our contrastive learning framework for ransomware detection, we define a hybrid loss function consisting of three major components:

\subsubsection{Contrastive Loss}  
Given a trace from an anchor program \( x^a \), a positive sample \( x^+ \) is selected either from another program within the same class or as a temporally modified version of \( x^a \). A negative sample \( x^- \) is drawn from a program of the opposite class. As introduced in Section~\ref{sec:ssl}, the goal is to bring positive pairs \( (x^a, x^+) \) closer in the feature space while pushing negative pairs \( (x^a, x^-) \) farther apart. Let \( h^a, h^+, h^- \) denote the hidden sequences encoded by the RNN-based encoder. The pairwise contrastive loss is defined as:
\begin{equation}
\mathcal{L}_{\text{pair}} = d(h^a, h^+) - d(h^a, h^-)
\end{equation}

\subsubsection{Intra-Class Clustering Loss}  
Unlike computer vision tasks, as illustrated in Figure~\ref{fig:contrastive_learning}, where two images of the same class often share obvious semantic features, ransomware and benign programs within the same class may exhibit significant behavioral diversity. To ensure consistency, we introduce a clustering loss that minimizes intra-class variance in the feature space.

Let \( \mu_k \) represent the centroid of class \( k \in \{0, 1\} \), where 0 and 1 correspond to benign and ransomware samples, respectively. Then:
\begin{equation}
\mathcal{L}_{\text{cluster}} = \sum_{h_i \in \mathcal{D}} \| h_i - \mu_{y_i} \|^2, \quad \text{where } \mu_k = \frac{1}{|\mathcal{D}_k|} \sum_{h_j \in \mathcal{D}_k} h_j
\end{equation}
Here, \( \mathcal{D}_k \) is the set of all encoded embeddings belonging to class \( k \), and \( y_i \) is the label for sample \( h_i \). 
This term reduces intra-class variation, improving class separability and aiding the downstream classification.
\subsubsection{Latency-Aware Loss}
To minimize detection delay, we further introduce a latency penalty that encourages early divergence in feature distance between benign and malicious traces. During training, for each sample pair $(x^a, x^-)$, we compute the earliest timestep $t_{\text{div}}$ at which the DTW cost exceeds a defined threshold $\delta$. Then we penalize longer detection latency as:
\begin{equation}
\mathcal{L}_{\text{latency}} = \frac{t_{\text{div}}}{T}
\end{equation}
where $T$ is the total sequence length. This term encourages the model to trigger a meaningful separation in the feature space as early as possible.

Finally, we combine the above components into a unified objective:
\begin{equation}
\mathcal{L}_{\text{total}} = \lambda_1 \mathcal{L}_{\text{pair}} + \lambda_2 \mathcal{L}_{\text{cluster}} + \lambda_3 \mathcal{L}_{\text{latency}}
\end{equation}
The coefficients $\lambda_1, \lambda_2, \lambda_3$ are hyperparameters balancing each components.

\subsection{NAS-guided Downstream Classifier}
\label{sec:nas}



    


The encoder outputs are passed to a downstream classifier that performs final ransomware detection. While existing approaches often rely on fixed, manually designed architectures, such classifiers may overfit to specific ransomware types and struggle to accommodate previously unseen variants.

To address this limitation, we adopt a \textit{Neural Architecture Search (NAS)} strategy~\cite{elsken2019neural} to automatically discover classifier structures that are both expressive and adaptive.

Specifically, our NAS process follows a one-shot search paradigm with two main phases:
\begin{itemize}
    \item \textbf{Supernet Construction:} We construct a multi-layer Supernet, where each layer includes multiple candidate operations (e.g., GRUs, fully connected layers, and non-linear activations). This serves as a universal search space of model architectures.

    \item \textbf{Pruning:} After training the Supernet, we apply gradient-based pruning~\cite{zullich2021speeding} to remove redundant or underperforming components, yielding a compact yet high-performing classifier architecture tailored to the current detection task.
\end{itemize}

Following the initial architecture search and pruning, our framework supports fast adaptation to emerging ransomware variants by performing lightweight retraining. Specifically, we fine-tune the selected components within the pre-trained Supernet without restarting the entire searching process, significantly reducing time and resource demands during retraining.

\begin{figure}[htbp]
\centering
\vspace{-0.1in}
\includegraphics[width = 0.5\textwidth]{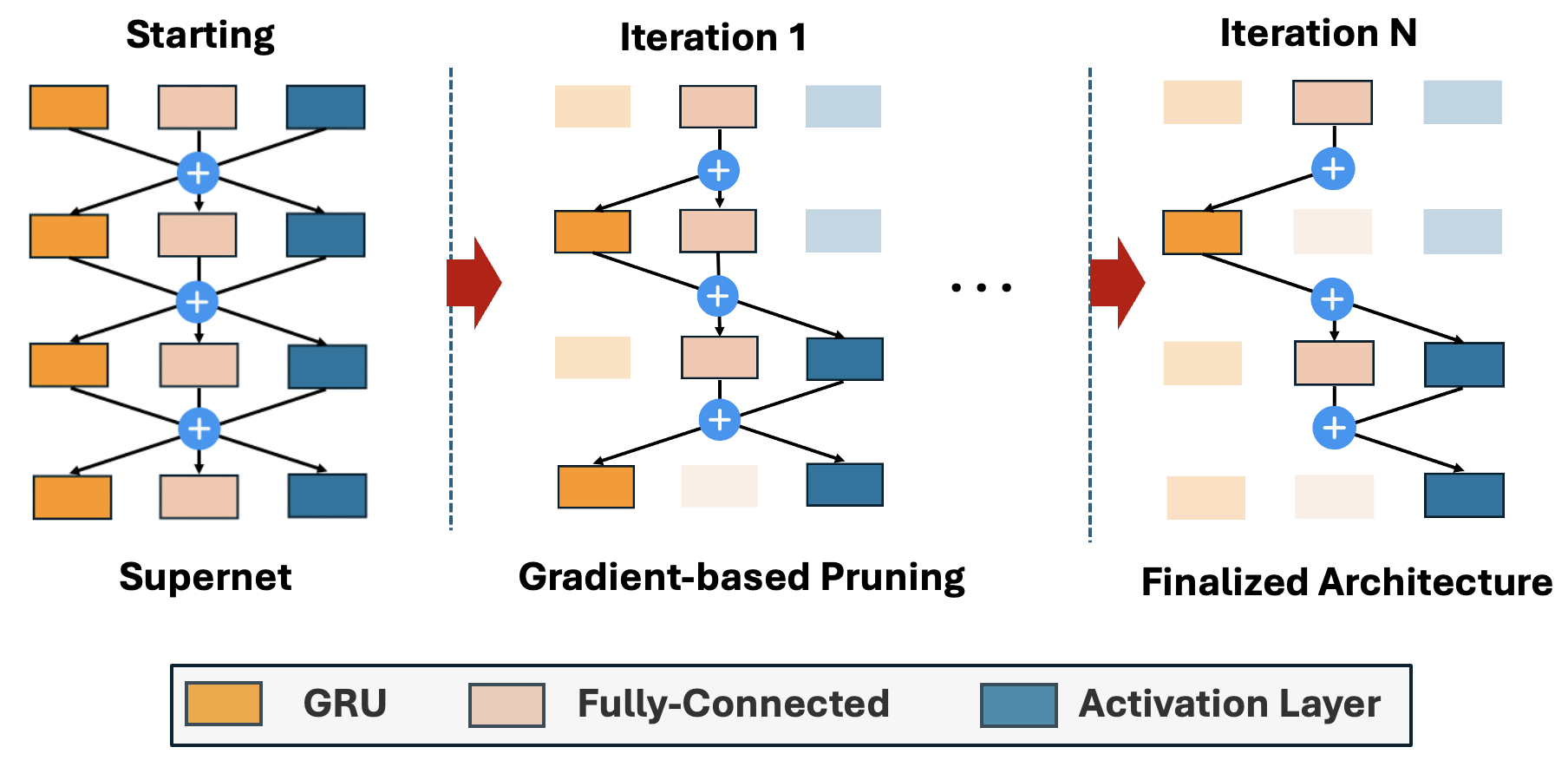}
\vspace{-0.1in}
\caption{One-shot NAS workflow for downstream classifier construction. Starting from a large Supernet, gradient-based pruning removes less important paths to form a lightweight streamlined architecture.}
\label{fig:oneshot}
\end{figure}

\subsection{Real-Time Detection and Rollback}
\label{sec:mitigation}

With the upstream encoder and downstream classifier, our proposed framework operates in a real-time detection loop. The runtime trace segments are continuously extracted in a window-based manner and are processed by the encoder to extract temporal embeddings. These embeddings are then passed to the NAS-optimized classifier, which raises an alert once ransomware is detected. 

However, ransomware may still partially encrypt user files before triggering an alert. To mitigate this, we incorporate a lightweight, system-level rollback mechanism that works along with the detection framework.

Specifically, during each sliding window, the system monitors accessed files and creates temporary backups using system built-in commands such as \texttt{rsync}. These backups are deleted if no threat is detected to reduce memory cost. If malicious activity is confirmed, the process is immediately terminated and the affected files are restored using the most recent backup, offering just-in-time mitigation to ransomware attacks.


\section{Experimental Evaluation}
\label{sec:exp}

In this section, we present a comprehensive experimental evaluation of the proposed method. We first introduce the experiment setup, followed by an overall assessment of detection accuracy and robustness against evasive techniques. Next, we demonstrate the feature extraction capability of the contrastive learning-based upstream encoder, followed by the evaluation of detection latency. Finally, we assess the adaptivity of the models to unseen variants, assessing the effectiveness of the NAS-optimized downstream classifier.



    
    
    

\subsection{Experiment Setup}

\noindent{\underline{\bf Benchmark:}} 
We conducted experiments on a Linux workstation to evaluate detection accuracy, robustness, latency, and adaptivity. Six ransomware variants,\textit{WannaCry}, \textit{Locky}, \textit{Cerber}, \textit{Vipasana}, \textit{Petya}, and \textit{Ryuk} were selected to enable comprehensive evaluation. Benign samples were collected from the SPEC CPU benchmark suite~\cite{SPEC}, system utilities, and common user applications. During execution, ETB logs were captured via UART at 50ms intervals. A window size of 500ms was used to segment traces, balancing responsiveness and system overhead. Totally, 2100 program traces were collected and evenly split between benign and malicious classes. The training and testing configurations are tailored to each evaluation objective, including detection accuracy, adversarial robustness, and adaptivity to unseen variants. A detailed description of the dataset split and task-specific setups is provided in their respective subsections.


\noindent{\underline{\bf Implementation:}} 
Our framework was implemented using PyTorch and executed on a workstation equipped with a 3.70GHz Intel Xeon CPU and 128 GB RAM. The upstream encoder was trained for 400 epochs using the Adam optimizer with a learning rate of 0.001. The downstream classifier was trained for 500 epochs with a dropout rate of 0.3 to prevent overfitting. We adopted an 80/20 train-validation split and applied cross-validation to reduce bias.

\noindent{\underline{\bf Approaches:}} 
We compared the following methods:

\begin{itemize}
    \item \textbf{SIA~\cite{baker2024static}:} A static-informed analysis approach using handcrafted signatures and entropy-based heuristics.
    
    \item \textbf{Ratafia~\cite{alam2019ratafia}:} A dynamic analysis method leveraging LSTM-based autoencoders for anomaly detection.
    
    \item \textbf{SCL~\cite{yang2025dynamic}:} A recently proposed supervised contrastive learning framework for ransomware detection.
    
    \item \textbf{Proposed:} The proposed method integrating contrastive learning and NAS .
\end{itemize}


\subsection{Case Study: Detection Accuracy}
\label{sec:exp_result}
\begin{table*}[htp]
\centering
\caption{Detection performance comparison on six ransomware variants.}
\label{tab:results}
\begin{tabular}{|c|cccc|cccc|cccc|cccc|}
\hline
\textbf{Benchmark} & \multicolumn{4}{c|}{SIA~\cite{baker2024static}} & \multicolumn{4}{c|}{Ratafia~\cite{alam2019ratafia}} & \multicolumn{4}{c|}{SCL~\cite{yang2025dynamic}} & \multicolumn{4}{c|}{Proposed} \\ \hline
\textbf{Ransomware} & \textbf{Acc} & \textbf{Prec} & \textbf{Rec} & \textbf{F1} & \textbf{Acc} & \textbf{Prec} & \textbf{Rec} & \textbf{F1} & \textbf{Acc} & \textbf{Prec} & \textbf{Rec} & \textbf{F1} & \textbf{Acc} & \textbf{Prec} & \textbf{Rec} & \textbf{F1} \\ \hline
WannaCry & 82.1 & 74.2 & 85.5 & 0.79 & 88.2 & 87.0 & 89.1 & 0.88 & 93.4 & 91.2 & 95.1 & 0.93 & 96.3 & 95.5 & 97.0 & 0.96 \\ \hline
Locky & 79.4 & 70.0 & 83.2 & 0.76 & 84.5 & 83.1 & 86.8 & 0.85 & 92.8 & 89.4 & 96.0 & 0.93 & 95.8 & 94.8 & 96.7 & 0.96 \\ \hline
Cerber & 76.7 & 67.1 & 81.5 & 0.73 & 86.9 & 84.5 & 89.8 & 0.87 & 85.1 & 82.3 & 88.4 & 0.85 & 95.0 & 93.5 & 96.1 & 0.95 \\ \hline
Vipasana & 75.8 & 65.4 & 80.2 & 0.72 & 83.6 & 82.0 & 85.7 & 0.84 & 77.2 & 70.5 & 84.3 & 0.77 & 95.5 & 94.0 & 96.8 & 0.95 \\ \hline
Petya & 84.3 & 75.1 & 87.9 & 0.81 & 89.0 & 87.3 & 90.6 & 0.89 & 92.0 & 90.8 & 92.2 & 0.91 & 96.7 & 95.9 & 97.5 & 0.97 \\ \hline
Ryuk & 80.5 & 69.3 & 84.6 & 0.76 & 85.5 & 83.6 & 87.2 & 0.85 & 90.2 & 88.0 & 91.5 & 0.89 & 95.9 & 94.6 & 97.1 & 0.96 \\ \hline
\textbf{Average} & \textbf{79.8} & \textbf{70.2} & \textbf{83.8} & \textbf{0.76} & \textbf{86.3} & \textbf{84.6} & \textbf{88.2} & \textbf{0.86} & \textbf{88.4} & \textbf{85.4} & \textbf{91.3} & \textbf{0.87} & \textbf{95.9} & \textbf{94.7} & \textbf{96.9} & \textbf{0.96} \\ \hline
\end{tabular}
\end{table*}

Table~\ref{tab:results} presents the detection performance of all evaluated approaches across six ransomware variants. The models are trained and tested with a random 80\%/20\% split over the entire dataset. We report four evaluation metrics, accuracy (Acc), precision (Pre), recall (Rec), and F1-score (F1), with each row corresponding to a specific ransomware family. 

As shown in the table, the static analysis based method SIA lags behind the other approaches, with an average accuracy of 79.8\%. Its reliance on static signatures leads to high false positive rates, and consequently, the lowest average precision (70.2\%). Ratafia as a dynamic analysis approach that applies autoencoders to runtime behavior, improves upon SIA with an average accuracy of 86.3\%, but it relies on manually crafted features, making it struggle to effectively capture subtle behavior transitions and long-range dependencies in ransomware activity, leading to suboptimal recall and F1 score. SCL achieves strong performance on certain variants (e.g., Cerber and Petya), but suffers from high variability. Its accuracy drops down to 77.2\% on Vipasana, resulting in a performance spread of over 16\%. This performance degradation can be attributed to Vipasana’s unique behavior of encrypting files offline, without contacting control servers, which deviates from typical ransomware dynamic behaviors.

In contrast, our {proposed method} consistently outperforms all baselines, achieving an average accuracy of 95.9\% and an F1-score of 0.96 across all benchmarks. We attribute the out-performance of the proposed method to the combination of contrastive self-supervised learning and NAS-guided architecture optimization (detailed further in Section~\ref{sec:expnas}).



\subsection{Case Study: Robustness}
\label{sec:exprobust}

We further evaluate the robustness of all methods against evasive attacks. In ransomware detection, robustness is a critical aspect, especially considering the adversarial nature of modern ransomware variants, which often employ evasion techniques to bypass conventional detection tools 

We generate evasive testing samples using three common strategies: (1) \textit{code morphing}, where semantically redundant instructions are injected to obscure recognizable patterns; (2) \textit{delayed activation}, in which encryption routines are shifted to later execution stages to avoid early detection; and (3) \textit{logic reordering}, where benign-looking code blocks are interleaved with malicious logic to disrupt temporal sequence integrity.

\begin{figure}[htbp]
\centering
\vspace{-0.1in}
\includegraphics[width = 0.42\textwidth]{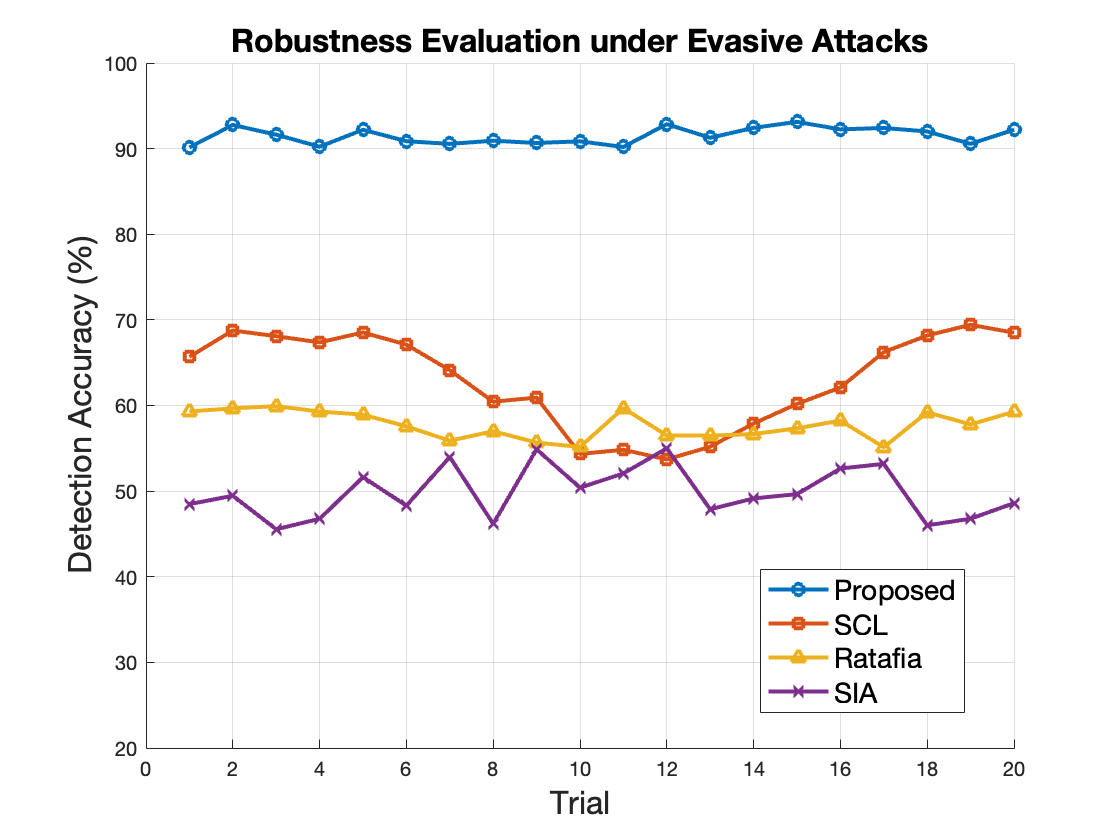}
\vspace{-0.1in}
\caption{Accuracies across 20 trials on evasive Ransomware attacks.}
\vspace{-0.1in}
\label{fig:robustness}
\end{figure}

Figure~\ref{fig:robustness} presents the accuracy of each method under evasive attacks. To enable fair comparison, we randomly apply different strategies with sampled traces and repeat the experiments for 20 trials. As shown, all three baseline approaches suffer notable performance degradation. {SIA}, due to its static signature reliance, fails to detect morphed or delayed variants. {Ratafia}'s performance also declines sharply, indicating vulnerability to timing manipulations. {SCL} exhibits slightly better robustness due to its feature learning capability, but still suffers from inconsistent performance under logic reordering.

In contrast, our proposed method maintains stable accuracy across all variants. We attribute this resilience to three core factors. First, our method avoids handcrafted features and instead learns representations through contrastive learning, making it less sensitive to surface-level obfuscations. Second, the use of hardware-assisted runtime monitoring ensures the core malicious behavior is still observed anyway. Additionally, our framework utilizes dynamic time warping (DTW) during feature extraction, which aligns temporally displaced behavior patterns, mitigating the reordering strategies.

\vspace{-0.03 in}
\subsection{Case Study: Feature Extraction}
\label{sec:expnas}

As discussed in Section~\ref{sec:ssl}, our proposed method leverages contrastive learning to eliminate the need for manually crafted features and enable more generalized feature extraction. To validate this claim, we compare the feature representations learned by our method against those extracted by Ratafia~\cite{alam2019ratafia}, which also uses RNNs for encoding but relies on manually defined feature types.

\begin{figure}[htbp]
\centering
\vspace{-0.15in}
\includegraphics[width = 0.5\textwidth]{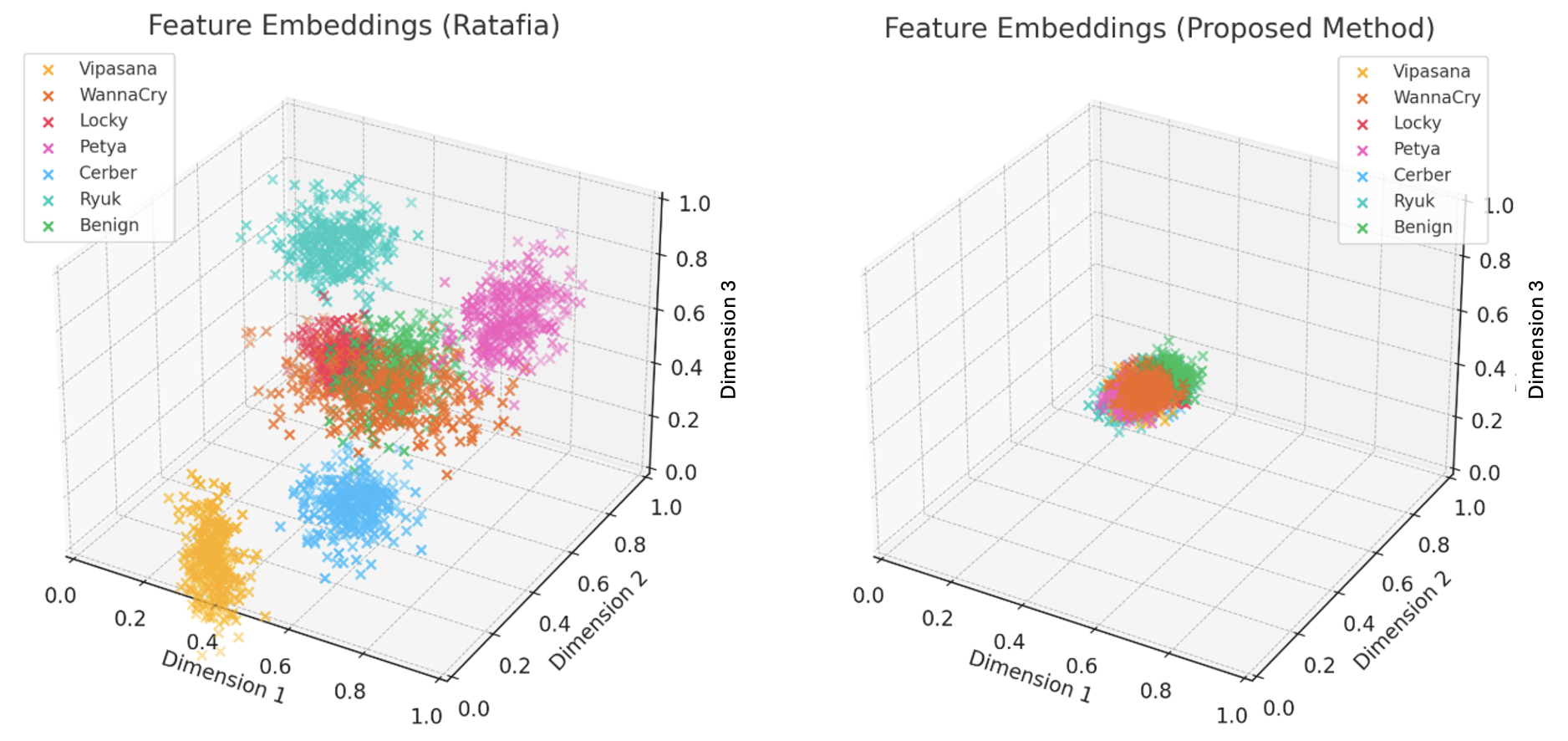}
\vspace{-0.2in}
\caption{Visualization of latent feature embeddings across ransomware variants using Ratafia (left) and the proposed contrastive learning method (right).}
\vspace{-0.15in}
\label{fig:feature_pca}
\end{figure}

We apply both methods to generate feature embeddings for all six ransomware variants, followed by the principal component analysis (PCA) to reduce the dimensionality to three for better visualization (Figure~\ref{fig:feature_pca}). As shown in the figure, in the case of Ratafia, several variants such as \textit{Ryuk} and \textit{Vipasana} appear loosely distributed and scattered away from the main clusters, making it difficult for a basic classifier to group these variants into a single category. This illustrates Ratafia’s limitation in capturing adaptive ransomware features across variants, which also explains its unstable classification performance discussed in Section~\ref{sec:exp_result}.

In contrast, the embeddings produced by our contrastive learning approach form a compact cluster for all six ransomware variants. This indicates that our method is effective in mapping ransomware traces close together in the feature space, regardless of their variant-specific characteristics. This well-structured representation facilitates the robustness of the framework, and explains the improved accuracy of our work demonstrated in earlier experiment results (Table~\ref{tab:results}).

\subsection{Case Study: Detection Latency}
\label{sec:latency}

\begin{figure}[htbp]
\centering
\vspace{-0.1in}
\includegraphics[width = 0.5\textwidth]{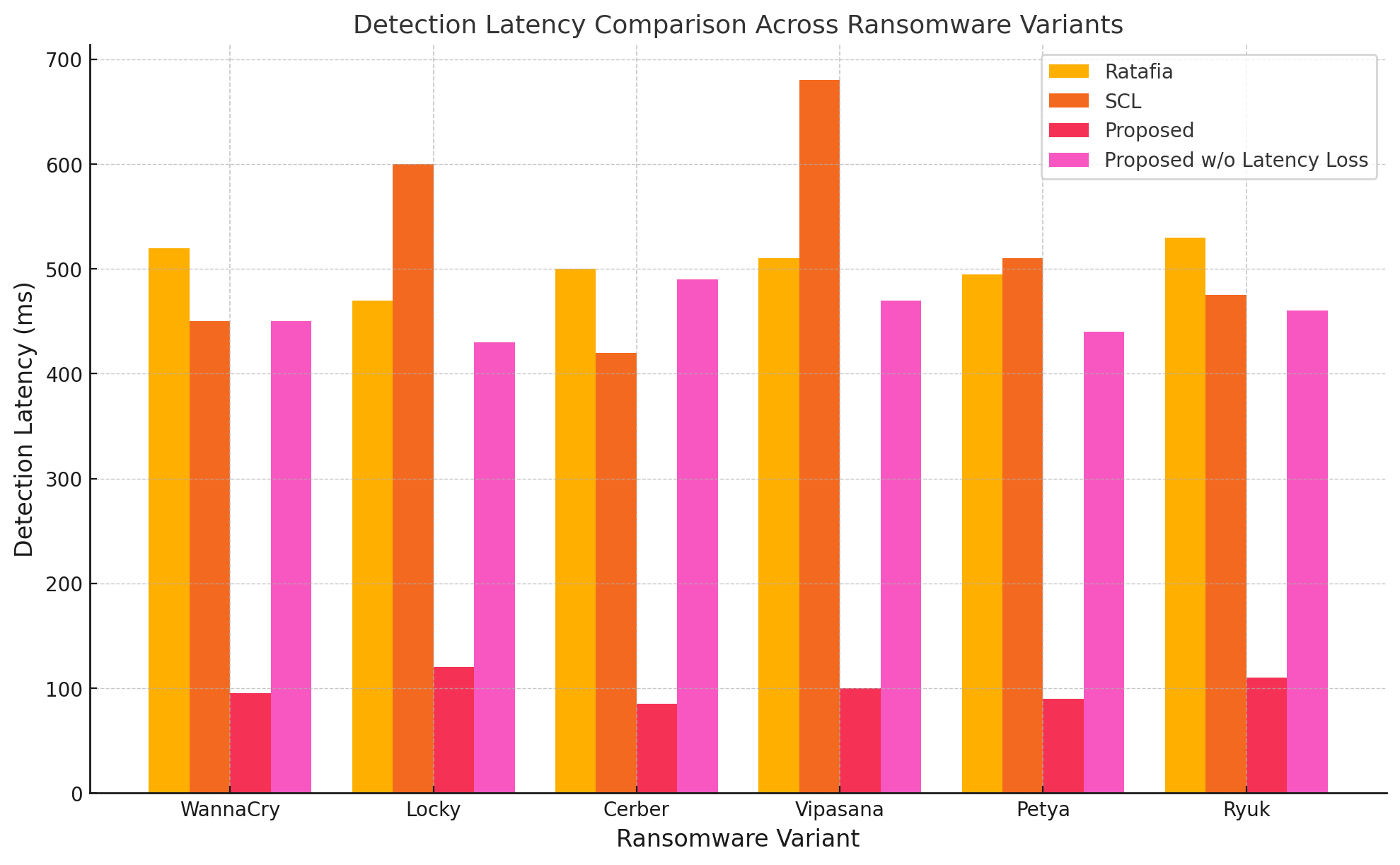}
\vspace{-0.2in}
\caption{Detection latency for different ransomware variants.}
\vspace{-0.2in}
\label{fig:latency}
\end{figure}

Detection latency is another critical factor in ransomware mitigation. Figure~\ref{fig:latency} presents the average detection latency of different methods across six ransomware variants. Since SIA is a static analysis method whose detection occurs before program execution, it is excluded from this comparison. Instead, we include an ablated version of our proposed model without the latency-aware loss to better illustrate its contribution.

Among all the evaluated methods, our proposed approach consistently achieves the lowest latency, with detection occurring in under 100 milliseconds on average. This demonstrates the model’s ability to raise timely alerts early in the infection process, enabling effective mitigation, as well as reducing the overhead for file backup operations introduced in Section~\ref{sec:mitigation}. We attribute this improvement to the inclusion of the latency-aware loss during training, as it explicitly encourages the model to reach earlier prediction. As shown in the figure, removing this component leads to a noticeable increase in latency (typically ranging from 400 to 500 milliseconds), confirming its essential role in reducing detection delay.

\subsection{Case Study: Adaptivity}
\label{sec:adapt}

We evaluate the adaptability of all methods from two key perspectives: (1) resilience to forgetting, and (2) retraining overhead when adapting to unseen ransomware variants.

To ensure a fair comparison, we simulate a transfer-learning scenario where each model is trained on three randomly selected ransomware families and then evaluated on the remaining unseen ones. Table~\ref{tab:adaptivity} reports four metrics: 
\begin{enumerate}
    \item Accuracy on unseen variants \textbf{before} retraining, 
    \item Accuracy on unseen variants \textbf{after} retraining, 
    \item Accuracy on previously seen variants \textbf{after} retraining (to measure forgetting), and 
    \item Average retraining time until convergence.
\end{enumerate}

All methods experience performance drops on unseen variants prior to retraining. While retraining improves performance, the first three baselines suffer from catastrophic forgetting, exhibiting degraded accuracy on previously seen samples. In contrast, our method maintains high accuracy on both seen and unseen variants after retraining, demonstrating strong resilience to forgetting. Additionally, our method achieves the shortest retraining time (79.8 seconds).

We attribute this efficiency to two factors: (1) the contrastive learning encoder is able to learn generalized representations from limited data, reducing total epochs to convergence for retraining,; and (2) the downstream classifier is instantiated from a pre-trained Supernet, requiring only lightweight parameter-tuning rather than full architectural redesign. 

\begin{table}[htp]
\vspace{-0.1 in}
\caption{Adaptivity Evaluation: Accuracy and retraining overhead.}
\vspace{-0.1 in}
\label{tab:adaptivity}
\begin{tabular}{|cc|c|c|c|c|}
\hline
\multicolumn{2}{|c|}{\textbf{Metric}}                           & \textbf{SIA} & \textbf{Ratafia} & \textbf{SCL} & \textbf{Proposed} \\ \hline
\multicolumn{1}{|c|}{\multirow{2}{*}{Pre-Retraining}}  & Seen   & 80.1\%       & 85.4\%           & 91.2\%       & 95.6\%            \\ \cline{2-6} 
\multicolumn{1}{|c|}{}                                 & Unseen & 63.4\%       & 70.5\%           & 76.2\%       & 81.0\%            \\ \hline
\multicolumn{1}{|c|}{\multirow{2}{*}{Post-Retraining}} & Seen   & 76.3\%       & 84.1\%           & 89.7\%       & 94.8\%            \\ \cline{2-6} 
\multicolumn{1}{|c|}{}                                 & Unseen & 70.2\%       & 78.4\%           & 84.6\%       & 94.1\%            \\ \hline
\multicolumn{2}{|c|}{Retraining Time (s)}                       & 274.5        & 1191.0           & 579.2        & 79.8              \\ \hline
\end{tabular}
\vspace{-0.1 in}
\end{table}

\subsection{Case Study: Overhead Analysis}


Table~\ref{tab:our_overhead} presents the overhead associated with the proposed method. As we can see, the proposed method maintains acceptable overhead for PC-level systems. While the NAS search process introduces a one-time cost (1.2 hours), this step is performed only during the initial design phase and does not impact runtime or future retraining for adaptive updates. As for inference overhead, the model remains highly efficient, with a total latency of 20.3 ms per sample and a memory footprint of 19.0 MB. These metrics suggest that the system can be deployed in real-time environments, including resource-constrained or endpoint devices.

\begin{table}[htbp]
\centering
\vspace{-0.1 in}
\caption{Breakdown of training and inference overhead for each model component.}
\vspace{-0.1 in}
\label{tab:our_overhead}
\begin{tabular}{lccc}
\toprule
\textbf{Metric} & \textbf{Encoder} & \textbf{Classifier} & \textbf{Total} \\
\midrule
\multicolumn{4}{l}{\textit{Training Overhead}} \\
Contrastive Pretraining Time (hrs) & 0.3 & --   & 0.3 \\
NAS Search Time (hrs)              & --  & 1.2  & 1.2 \\
Retraining Time (s)                & 20.5 & 59.3 & \textbf{79.8} \\
Model Size (M parameters)          & 2.4 & 1.1  & 3.5 \\
\midrule
\multicolumn{4}{l}{\textit{Inference Overhead}} \\
Latency (ms/sample)                & 13.1 & 7.2  & \textbf{20.3} \\
Memory Footprint (MB)              & 11.9 & 7.1 & \textbf{19.0} \\
\bottomrule
\end{tabular}
\vspace{-0.1 in}
\end{table}

\section{Conclusion}
\label{sec:conc}

Ransomware remains a critical cybersecurity threat due to its rapid evolution, irreversible damage, and evasive behavior. In this work, we propose a real-time ransomware detection that integrates contrastive learning and neural architecture search (NAS). By leveraging hardware-assisted runtime monitoring and dynamic time warping, our approach eliminates the need for handcrafted features while offering robustness against evasion techniques. A latency-aware loss is also applied to significantly reduce detection latency. Additionally, the utilization of NAS ensures our framework's strong adaptability to unseen ransomware variants with minimal retraining overhead. Experimental results confirm that our method outperforms existing approaches in both detection accuracy and response time while maintaining resilience against evasive techniques.

\bibliographystyle{IEEEtran}
\bibliography{bibliography}

\end{document}